%% file: AR-GENERIC-ARXIV.tex
\documentclass[10pt]{article}
\usepackage[T1]{fontenc}
\usepackage[utf8]{inputenc}
\usepackage{authblk}
\usepackage{arxiv}

\input{header.tex}

\title{Thermodynamics-informed neural networks for physically realistic mixed reality}

\author[1]{Quercus Hern\'andez}
\author[2]{Alberto Bad\'ias}
\author[3,4]{Francisco Chinesta}
\author[1]{El\'ias Cueto}
\affil[1]{{\small Aragon Institute of Engineering Research (I3A). University of Zaragoza. Zaragoza, Spain.}}
\affil[2]{{\small Higher Technical School of Industrial Engineering, Polytechnic University of Madrid. Madrid, Spain.}}
\affil[3]{{\small ESI Group chair. PIMM Lab. ENSAM Institute of Technology. Paris, France.}}
\affil[4]{{\small CNRS@CREATE LTD. Singapore.}}

\begin{document} 

\maketitle

\begin{abstract}
The imminent impact of immersive technologies in society urges for active research in real-time and interactive physics simulation for virtual worlds to be realistic. In this context, realistic means to be compliant to the laws of physics. In this paper we present a method for computing the dynamic response of (possibly non-linear and dissipative) deformable objects induced by real-time user interactions in mixed reality using deep learning. The graph-based architecture of the method ensures the thermodynamic consistency of the predictions, whereas the visualization pipeline allows a natural and realistic user experience. Two examples of virtual solids interacting with virtual or physical solids in mixed reality scenarios are provided to prove the performance of the method.
\end{abstract}

\section{Introduction}\label{sec:intro}

Computer science advances in the last decades led us to experience in a relatively short lapse of time three major technological innovations: the personal computer, the Internet, and mobile devices. Currently, we are at the beginning of a fourth paradigm of computing innovations involving immersive technologies such as Virtual Reality (VR), Augmented Reality (AR) or Mixed Reality (MR). All this is possible due to huge advances in machine learning techniques and hardware improvements applied to computer graphics and computer vision. 

It is clear that this new paradigm seeks to revolutionize technology in the next years and will have a great impact in society such as smart cities \cite{allam2022metaverse, veeraiah2022enhancement}, new teaching methods \cite{rospigliosi2022metaverse} or economic paradigms \cite{wang2022metasocieties}. Technology companies have already created numerous digital platforms, such as the Metaverse \cite{mystakidis2022metaverse,kraus2022facebook} or the Omniverse \cite{hummel2019leveraging,li2022exploring}, in order to develop their own immersive technologies. 

In most of the cases, virtual worlds are required to resemble our real world as much as possible, so many disciplines come into play (traditionally, computer graphics and computer vision). However, virtual worlds need to be dynamic rather than static so the user can responsively interact with a changing virtual world. From that perspective, physics simulation plays a major role and it is required to be real-time. On the one hand, current real-time physic engines rely on severe simplifications of the governing dynamical equations and are limited to very simple material models and constitutive phenomena. On the other hand, classical engineering methods for solid and fluid simulations, such as the finite element or finite differences methods, have the consistency of decades of theoretical research in terms of the convergence to a consistent physical solution but are too expensive to achieve real-time framerates. However, these last methods can be used to generate a rich and consistent database to train a fast AI accelerated by the recent advances in machine learning procedures.

In this work, we aim to merge the physical consistency of classical simulation methods with the speed of real-time physics engines using a deep learning approach to develop a real-time interactive simulator. Although the formulation is general for a wide variety of dynamical systems, in this work we focus on nonlinear solid mechanics. The results are consistent with the laws of thermodynamics by construction and are able to achieve real-time performance in general load cases which were not previously seen by the network. In \mysecref{sec:related}, we explore some of the related work involving real-time reality simulators and physics-informed deep learning. In \mysecref{sec:methods} we introduce the thermodynamically-informed graph neural networks together with the vision and visualization system used to record the demo videos. \mysecref{sec:results} shows the application of the proposed algorithm in two different solids together with error plots, and future work, limitations and conclusions are provided in \mysecref{sec:conc}.

\section{Related work}\label{sec:related}

Real-time physics engines such as PhysX \cite{maciel2009using,d2013physx} or Havok \cite{wang2015method} have mostly been developed in the videogame industry. Even if those engines allow to program custom dynamical models, they usually rely on simplified mass-spring models or rigid body dynamics to achieve high framerates in modern videogames. Multiple research lines remain open trying to leverage the physical consistency of the results with low computational requirements.

\subsection{Model order reduction-based simulators}

Several authors solved the mentioned problems by creating a reduced order model of the system. These methods consist of a two-stage procedure: an offline phase where the solution space is precomputed in a compressed representation, lying on a reduced-order manifold, and an online phase where the solution can be evaluated in real-time. The difference between each method lies on the specific projection technique used to compute the reduced manifold. 

Classical linear methods such as Principal Component Analysis \cite{berkooz1993proper,rama2016real} or reduced basis \cite{prud2002reliable, manzoni2015reduced} are fast and simple to implement but fail to capture more complex nonlinear phenomena. This inconvenience can be solved using nonlinear methods such as kernel-Principal Component Analysis \cite{scholkopf1997kernel}, Locally Linear Embedding \cite{roweis2000nonlinear,moya2019learning} or Proper Generalized Decomposition \cite{badias2017local,badias2020real}. Those techniques have similar disadvantages: as the solution is already precomputed, they are unable to handle different mesh discretizations and fail to generalize to unseen configurations.

\subsection{Deep learning based simulators}

The use of deep learning in real-time simulations has been widely explored in recent manuscripts, using neural networks as powerful function approximators with fast evaluation performance. The spirit is similar to the reduced order modelling: in the offline phase a neural network is trained with a set of examples and in the online phase the network can be fast evaluated in unseen scenarios.

For instance, several works avoid the real-time restrictions by decreasing the dimensionality of the problem by using autoencoders \cite{fulton2019latent,chen2022crom} in a similar fashion to the reduced order modelling methods. Other approaches are based on standard multilayer perceptrons, used as a collocation method for residual minimization \cite{fresca2022deep,odot2022deepphysics} or specific formulations for contact mechanics \cite{romero2022embodied,romero2022contact}. Those approaches require the prior knowledge of the governing equations. In the field of physics-informed machine learning, many methods have recently proposed to use neural networks such as solvers of partial differential equations (PDE) residuals in the context of general physics \cite{raissi2018hidden,raissi2019physics} or fluid mechanics \cite{eivazi2022physics,eivazi2022bphysics,cai2022physics}, identification and simulation of the dynamics of complex systems \cite{sanchez2020learning} or structure-preserving algorithms \cite{sanchez2019hamiltonian,chen2019symplectic,hernandez2021structure} with promising results. However, none of the mentioned methods have been implemented nor tested in an augmented or virtual reality setup, to the best of our knowledge.

This work presents a method to simulate real-time dynamics of one or more virtual systems interacting with physical objects in a mixed reality application. We require no governing equation, as it is learned from data using a thermodynamics-based formulation for non-equilibrium dynamical systems together with geometric deep learning. Furthermore, the trained graph neural network achieves real-time performance which enables a smooth user experience in the interaction of several virtual objects.

\section{Methodology}\label{sec:methods}

\subsection{Problem Statement}

The present work focuses on the deformation of virtual solid objects. Thus, we use the dynamical equilibrium equation of non-linear solid mechanics which balances the external and internal body forces with the acceleration of the solid. This is,
\begin{equation}\label{eq:equilibrium}
\bs{\nabla{P}}+\bs{B}= \rho \ddot{\bs u} \: \: \text{in} \: \Omega_{0},
\end{equation}
where $\bs B$ represents the volumetric force applied to the body and $\bs P$ the first Piola-Kirchhoff stress tensor. $\Omega_0 = \Omega(t=0)$ represents the undeformed configuration of the virtual solid. The solution is subjected to apropriate boundary conditions
\begin{align}
\begin{split}\nonumber
\bs{u}(\bs{X})=\overline{\bs{u}}\:\: &\text{on}\:\:\Gamma_{u},\\
\bs{PN}=\overline{\bs{t}}\:\: &\text{on}\:\:\Gamma_{t},
\end{split}
\end{align}
with $\Gamma_u$ and $\Gamma_t$ representing the essential (Dirichlet) and natural (Neumann) portions of the boundary $\Gamma=\partial \Omega$ of the solid. $\bs X$ is the undeformed position, $\bs N$ is the unit vector normal to $\Gamma=\partial \Omega_0$ and $\bs{\bar{u}}$, $\bs{\bar{t}}$ are the applied displacement and traction respectively. To complete the problem, some relationship between kinematic variables (displacements, strain) and dynamic variables (stresses) must be assumed. The constitutive equation is here chosen to be hyperelastic, with a strain energy function per unit volume $\Psi$ defined such that
\begin{equation}
\bs{S}=\dpar{\Psi}{\bs{E}}
\end{equation}
where $\bs{S}$ is the second Piola-Kirchhoff and $\bs{E}$ is the Green-Lagrange strain tensor. Viscoelastic effects are also considered using variable shear relaxation modulus via Prony series. The objective of the method is to solve \myeqref{eq:equilibrium} in a real-time interactive interface with a physics-based neural network trained with high fidelity solutions.

\subsection{Thermodynamics-informed graph neural networks}

We use a novel deep learning method \cite{hernandez2022thermodynamics} which aims to learn the dynamical evolution of a physical system using a graph-based approach. Its objective is to learn not the outcome of a given simulation under different conditions such as forces or boundary conditions, but the actual {\em physics} taking place, such that the learned simulator is not sensitive to changes in the mesh, for instance.

\begin{figure}[h]
\centerline{\includegraphics[width=\linewidth]{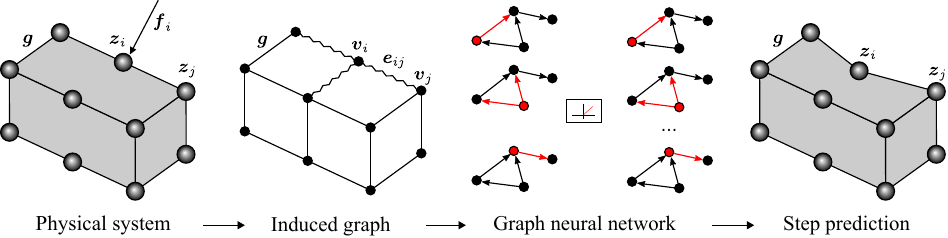}}
\caption{Thermodynamics-informed graph neural network architecture. The system is described as a set of state variables $\bs{z}_i$, global simulations parameters $\bs{g}$ and external boundary conditions $\bs{f}_i$. A graph $\mathcal{G}$ is constructed from the information of the physical system, defining vertex features $\bs{v}_i$, edge features $\bs{e}_{ij}$ and global features $\bs{g}$. The graph features are processed with a message-passing graph neural network. The step prediction is performed using the GENERIC integration scheme in \myeqref{eq:GENERIC}, which is repeated iteratively to get the complete rollout of the simulation.\label{fig:tignn}}
\end{figure}

The graph neural network architecture is thus constructed on top of a graph structure $\mathcal{G}=(V,\mathcal{E},\bs{g})$, where $V=\{1,...,n\}$ is a set of $\lvert V\rvert=n$ vertices, $\mathcal{E}\subseteq V\times V$ is a set of $\lvert \mathcal{E}\rvert=e$ edges and $\bs{g}\in \mathbf{R}^{F_g}$ is the global feature vector. Each vertex and edge in the graph is associated with a node in the finite element model from which data are obtained. The global feature vector defines the properties shared by all the nodes in the graph, such as constitutive properties.

To ensure translational invariance of the learned model, the position variables of the system $\bs{q}_i$, are assigned to the edge feature vector $\bs{e}_{ij}\in \mathbb{R}^{F_e}$ so the edge features represent relative distances ($\bs{q}_{ij}=\bs{q}_i-\bs{q}_j$) between nodes. The rest of the state variables are assigned to the node feature vector $\bs{v}_{i}\in \mathbb{R}^{F_v}$, while the external forces are included in a vector $\bs{f}_i\in \mathbb{R}^{F_f}$. We employ an encode-process-decode scheme \cite{battaglia2018relational}, built upon multilayer perceptrons (MLPs) shared between all the nodes and edges of the graph.

We use this architecture to learn the GENERIC structure of the evolution in time of the variables governing the virtual system \cite{ottinger1997dynamics, grmela1997dynamics}. It consists on splitting the system into a conservative and dissipative contribution. The conservative dynamics are defined using an energy potential $E$ and a symplectic Poisson matrix $\bs{L}$, which recover the Hamiltonian formalism, whereas the dissipative dynamics are described by the entropy potential $S$ and the dissipative or friction matrix $\bs{M}$, which accounts for the non-reversible dynamics. The time evolution of the state variables of the system $\bs{z}$ is described by the following equation
\begin{equation}\label{eq:GENERIC}
\frac{d\bs{z}}{dt}=\bs{L}\nabla E+\bs{M}\nabla S.
\end{equation}
By enforcing the so called degeneracy conditions $\bs{L}\nabla S=\bs{M}\nabla E=\bs{0}$ together with the algebraic properties of the $\bs{L}$ (skew-symmetric) and $\bs{M}$ (symmetric and positive semidefinite) matrices we ensure the energy conservation and the entropy inequality. Thus, we guarantee the thermodynamical consistency of the predictions. The neural network learns the parametrization of the Poisson and friction operators in lower triangular matrices, $\boldsymbol{l}$ and $\boldsymbol{m}$ respectively, and the energy and entropy potentials, $E$ and $S$. The GENERIC operators are then assembled as $\boldsymbol{L}=\boldsymbol{l}-\boldsymbol{l}^\top$ and $\boldsymbol{M}=\boldsymbol{m}\boldsymbol{m}^\top$, which enforces the skew-symmetry and positive semi-definiteness of the operators respectively. A scheme of the algorithm is presented in \myfigref{fig:tignn}.

We assume our virtual solids to be viscous-hyperelastic, so that the state variables for the proper description of their evolution in terms of the GENERIC formalism are the position $\bs{q}$, velocity $\bs{v}$ and the stress tensor $\bs{\sigma}$,
\begin{equation}\label{eq:beam_z}
\mathcal{S}=\{\bs{z}=(\bs{q},\bs{v},\bs{\sigma})\in\mathbb{R}^3\times\mathbb{R}^3\times\mathbb{R}^6\}.
\end{equation}
The edge feature vector contains the relative deformed position between nodes, to give a distance-based attentional flavour to the graph processing blocks and translational invariance. The velocity and strees tensor components are part of the node feature vector, concatenated to a two-dimensional one-hot vector $\bs{n}$ which represent the encastre and solid nodes respectively. The external load vector $\bs{f}_i$ is included in the node processor MLP as an external interaction. No global feature vector is needed in this case, resulting in the following feature vectors:
\begin{equation}\label{eq:beam_feat}
\bs{e}_{ij}=(\bs{q}_i-\bs{q}_j,\Vert\bs{q}_{ij}\Vert_2),\quad\bs{v}_i=(\bs{v},\bs{\sigma},\bs{n}).
\end{equation}
Thus, the dimensions of the feature vectors are $F_e=4$, $F_v=11$, $F_f=3$ and $F_g=0$.

\subsection{Vision system}

For an augmented or mixed reality application, we need to include virtual objects in a real scene. For that, it is necessary to have a sensor able to get information about the physical environment and a screen device to plot the resulting image. In the present work, we use a ZED Mini stereo vision system from Stereolabs, which is able to retrieve both a depth and RGB image of the captured snapshot. We plot the resulting real-time video stream in a computer screen, but could be extended to a VR headset or AR glasses.

\begin{figure}[h]
\centerline{\includegraphics[width=\linewidth]{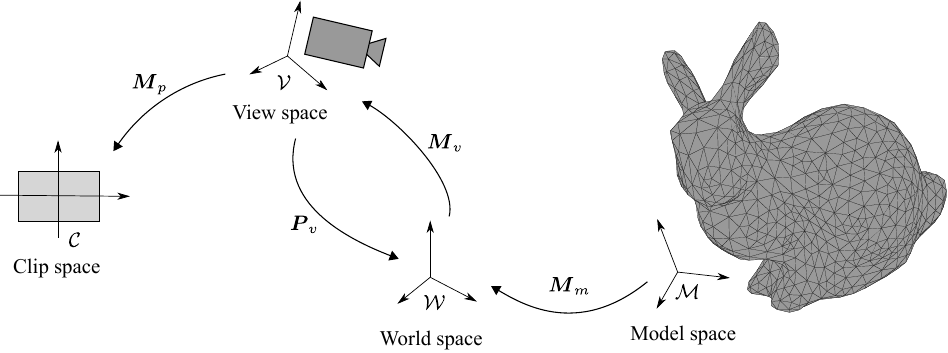}}
\caption{Model-view-projection transformation between the model and the clip coordinates. The dynamics and interactions are computed in the model space as 3D coordinates, whereas the visualization pipeline requires a 2D clipping coordinates in a fixed range between -1 and 1.\label{fig:mvp}}
\end{figure}

As the laws of motion are independent with respect to any frame of reference (objectivity), we use the model space to compute the dynamical state variables of the system. However, the visualization pipeline requires the coordinates to be in a normalized range of $\mathcal{C}=[-1,1]\times[-1,1]$ called the clip space. The transformation between the 3D model space and the 2D clip space is achieved by the definition of the Model-View-Projection matrix (see \myfigref{fig:mvp}). From now, the point coordinates are supposed to be in homogeneous coordinates.

\begin{itemize}
\item Model: The dynamics are computed in a local frame of reference called the model space $\mathcal{M}$. The position, orientation and scale of the model can be defined as a set of transformation matrices $\bs{T}_m$, $\bs{R}_m$ and $\bs{s}_m$ respectively with respect to the world space $\mathcal{W}$, considered to be the usual Euclidean space $\mathbb{R}^3$.  
\begin{equation}
\bs{x}_{\text{world}}=\bs{M}_m \bs{x}_{\text{model}}=\bs{T}_m\bs{R}_m\bs{s}_m\bs{x}_{\text{model}}.
\end{equation}

\item View: In a similar fashion, the viewing camera also has a model matrix defining its position in the world space, which is commonly designed as the extrinsic parameters $\bs{P}_v$ of the camera pose. This information can be obtained using a monocular pose estimator or triangulating with a stereo vision system. Thus, the camera or view space $\mathcal{V}$ can be determined by a set of transformations from the world space using the camera extrinsic parameters given by a rotation $\bs{R}_v$ and translation $\bs{T}_v$ matrices.
\begin{equation}
\bs{x}_{\text{view}}=\bs{M}_v \bs{x}_{\text{world}}={\bs{P}_v}^{-1}\bs{x}_{\text{world}}=(\bs{T}_v\bs{R}_v)^{-1}\bs{x}_{\text{world}}.
\end{equation}

\item Projection: The last transformation is in charge of projecting the 3D coordinates into the 2D clip space $\mathcal{C}$. First, in order to get a realistic visualization, we use a perspective projection $\bs{M}_p$ based on the camera viewing frustrum
\begin{equation}
\bs{M}_p=\begin{pmatrix}
\cot{\frac{\alpha}{2}} & 0 & 0 & 0\\
0 & \cot{\frac{\alpha}{2}} & 0 & 0\\
0 & 0 & -\frac{z_\text{far}+z_\text{near}}{z_\text{far}-z_\text{near}} & -\frac{2z_\text{far}z_\text{near}}{z_\text{far}-z_\text{near}}\\
0 & 0 & -1 & 0
\end{pmatrix},
\end{equation}
where $\alpha$ is the field of view of the camera and $z_\text{near}$ and $z_\text{far}$ are the minimum and maximum distance of the clipping plane. This creates a normalized 3D viewing box of the camera, which is then projected in the 2D clip space by the vertex shader. Thus, the final coordinates can be computed using the following equation:
\begin{equation}\label{eq:mvp}
\bs{x}_{\text{clip}}=\bs{M}_p \bs{x}_{\text{view}}=\bs{M}_p\bs{M}_v\bs{M}_m \bs{x}_{\text{model}}.
\end{equation}

\end{itemize}

By defining the Model-View-Projection matrix as $\bs{M}_p\bs{M}_v\bs{M}_m$ we can compute the clip coordinates directly from the model coordinates computed by the thermodynamics-informed neural network. 

\subsection{Visualization system}

The visualization of the resulting image is performed using OpenGL and the GLU library. This pipeline requires the definition of two spatial functions, run sequentially on the GPU, called the vertex and fragment shaders. The vertex shader defines the vertex position of the entities to display whereas the fragment (or texture) shader define the RGBA colors for each rasterized pixel.

Each vertex position is computed using \myeqref{eq:mvp} from the deformed configuration coordinates and the polygons are drawn based on the connectivity matrix of each solid. The color of each vertex is computed directly from the neural network prediction using a fixed colormap. Additionally, a basic lighting model was added to the fragment shader to increase the realism of the virtual objects using the Phong shading. This model computes each RGB pixel intensity as
\begin{equation}
\bs{I}=k_a\bs{i}_a+k_d(\bs{l}\cdot\bs{n})\bs{i}_d+k_s(\bs{r}\cdot\bs{v})^\beta \bs{i}_s,
\end{equation}
where $k_a$, $k_d$ and $k_s$ are the ambient, diffuse and specular reflection constants, $\beta$ is the shininess constant of the material and $\bs{i}_a$, $\bs{i}_d$ and $\bs{i}_s$ are the RGB color intensities of the ambient, diffuse and specular components. The illumination varies depending on the geometry of the scene and camera position, where $\bs{n}$ is the surface normal vector, $\bs{l}$ and $\bs{r}$ are the directions of the light source and its perfect reflection, $\bs{v}$ is the viewing direction and ``$\cdot$" is the dot product.

We have also implemented a depth or $z$-buffer in the fragment shader which compares on each pixel the depth of virtual and real objects on the scene and renders only the closest to the camera. This accounts for every occlusion that the vision system may encounter. The depth of the scene seen by the camera is computed directly using the stereo vision system.

\subsection{Collision and contact}

We use a high-fidelity hand tracker from MediaPipe \cite{zhang2020mediapipe} which provides an accurate localization of the finger tips. By using the inverse transformation of \myeqref{eq:mvp}, we can compute the 3D coordinates of the finger tips in the model space and compute the distance to each node of the virtual object. When this distance is less than a small threshold, collision is detected and a prescribed force is applied to the model. A similar procedure is applied to the collision of several virtual objects.

The code is implemented in Python using the PyOpenGL wrapper for the visualization and Pytorch Geometric \cite{fey2019fast} for the deep learning training and evaluation. The videos are generated using a standard desktop computer with a single Nvidia RTX2070 GPU and provided as supplementary material. The code for the implementation of the thermodynamics-informed neural networks and the video sequences are publicly available at \url{https://github.com/quercushernandez}.

\section{Results}\label{sec:results}

\subsection{Bending beams}\label{sec:beams}

\begin{figure}[h]
\centering
\includegraphics[width=\textwidth]{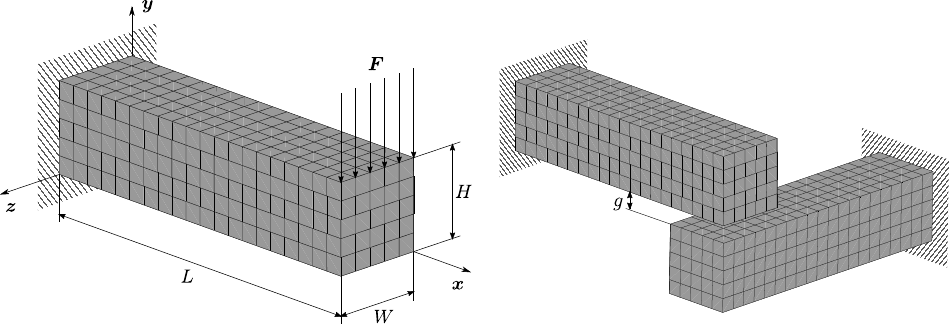}
\caption{Interacting beams geometry scheme. Both beams are angled with a gap between both, enabling their interaction.\label{fig:beams}}
\end{figure}

The first system is composed of two interacting viscoelastic beams. Both identical beams are assembled with a small gap between each other, allowing for a contact interaction, as depicted in \myfigref{fig:beams}.

The dimensions of the beams are $H=10$, $W=10$ and $L=40$. The finite element mesh from which data are obtained consists of $N_e=500$ hexahedral linear brick elements and $N=756$ nodes. The constitutive parameters of the hyperelastic strain energy potential are $C_{10}=1.5\nexp{5}$, $C_{01}=5\nexp{3}$, $D_1=10^{-7}$ and $\bar{g}_1=0.3$, $\bar{g}_2=0.49$, $\tau_1=0.2$, $\tau_2=0.5$ respectively for the two-term Prony series. A distributed load of $F=10^5$ is applied in 52 different positions with a perpendicular direction to the solid surface. Each simulation is composed of $N_T=20$ time increments of $\Delta t=5\cdot 10^{-2}$ s. Both beams are assembled in $90^{\circ}$ with a gap of $g=10$.

The graph neural network vertex and edge MLPs have two layers of $F_h=100$ neurons each, with 10 message passing sequential blocks. The training was performed for $N_\text{epoch}=1800$ epochs and learning rate $l_r=10^{-4}$.

\begin{figure}[!h]
  \begin{subfigure}[t]{.49\textwidth}
    \centering
    \includegraphics[width=\textwidth]{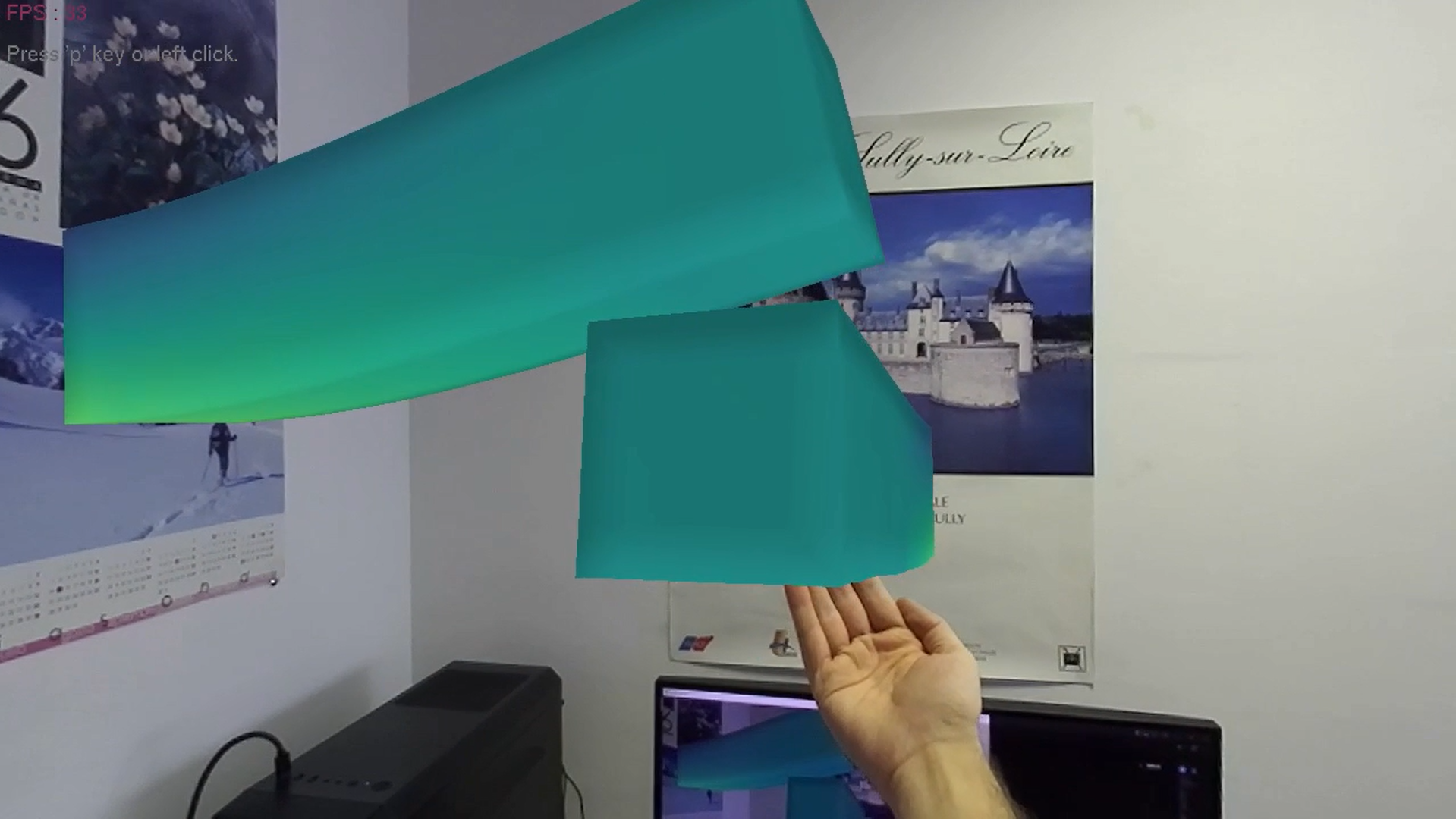}
  \end{subfigure}
  \hfill
  \begin{subfigure}[t]{.49\textwidth}
    \centering
    \includegraphics[width=\textwidth]{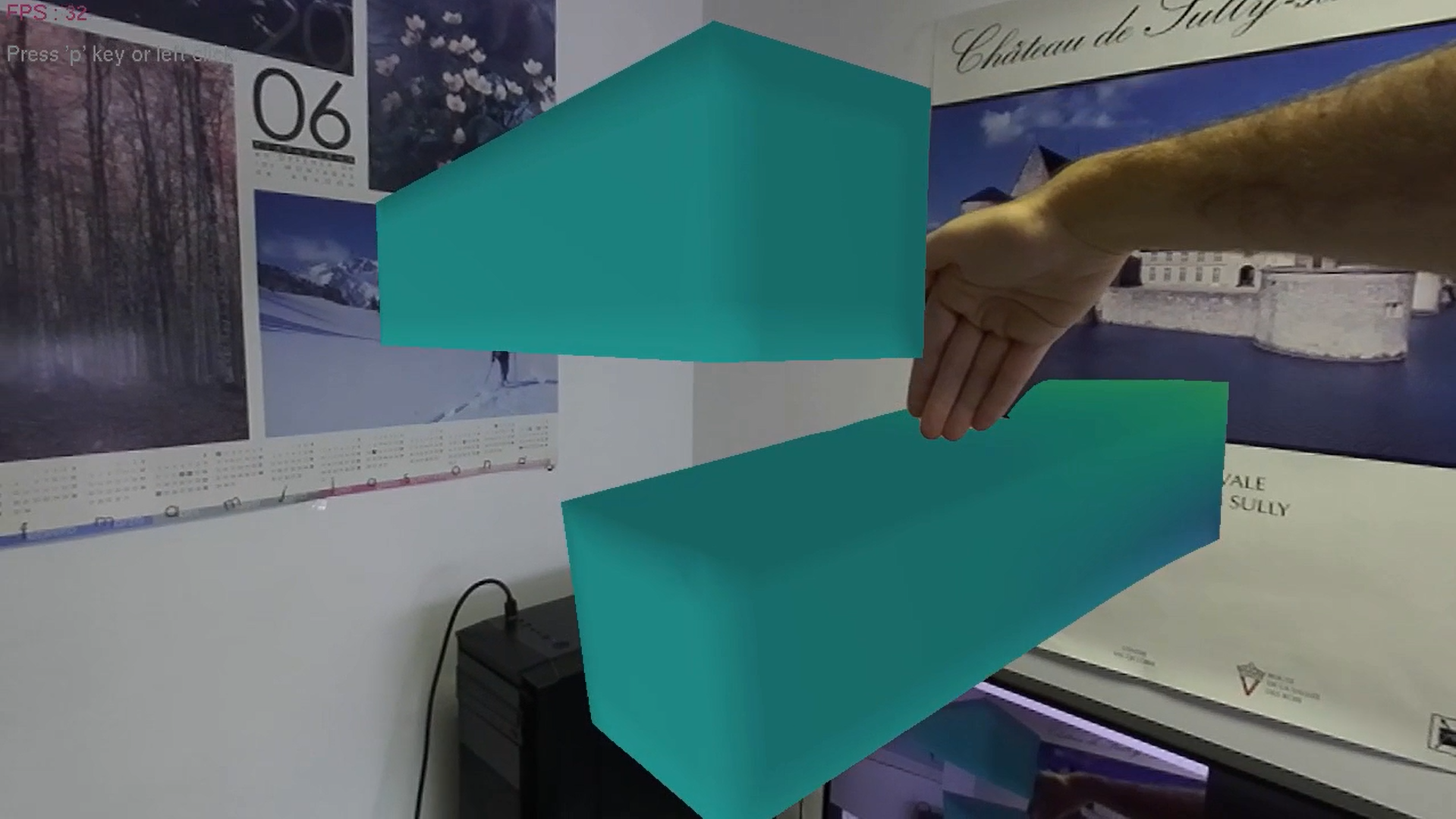}
  \end{subfigure}

  \medskip

  \begin{subfigure}[t]{.49\textwidth}
    \centering
    \includegraphics[width=\textwidth]{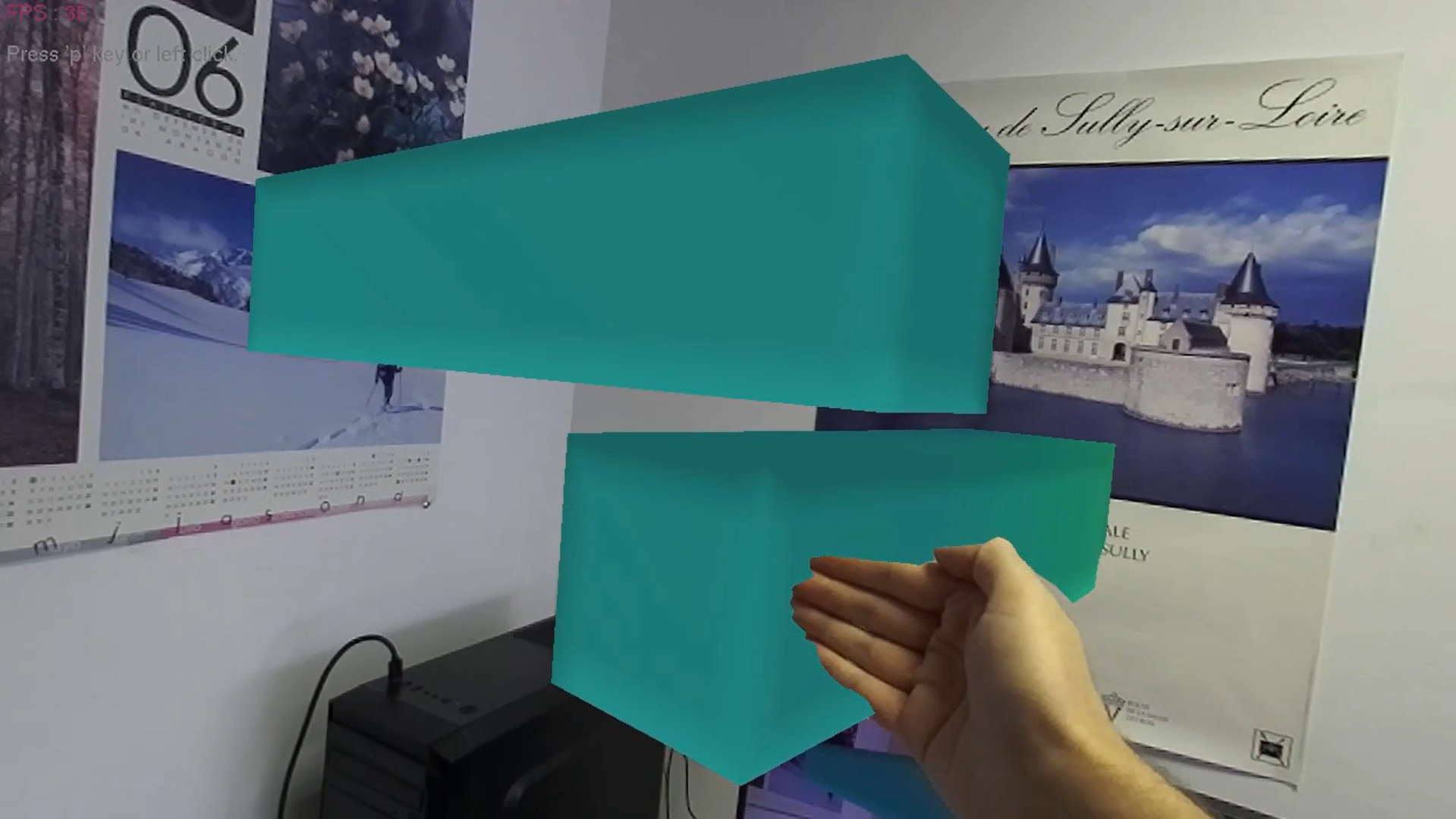}
  \end{subfigure}
  \hfill
  \begin{subfigure}[t]{.49\textwidth}
    \centering
    \includegraphics[width=\textwidth]{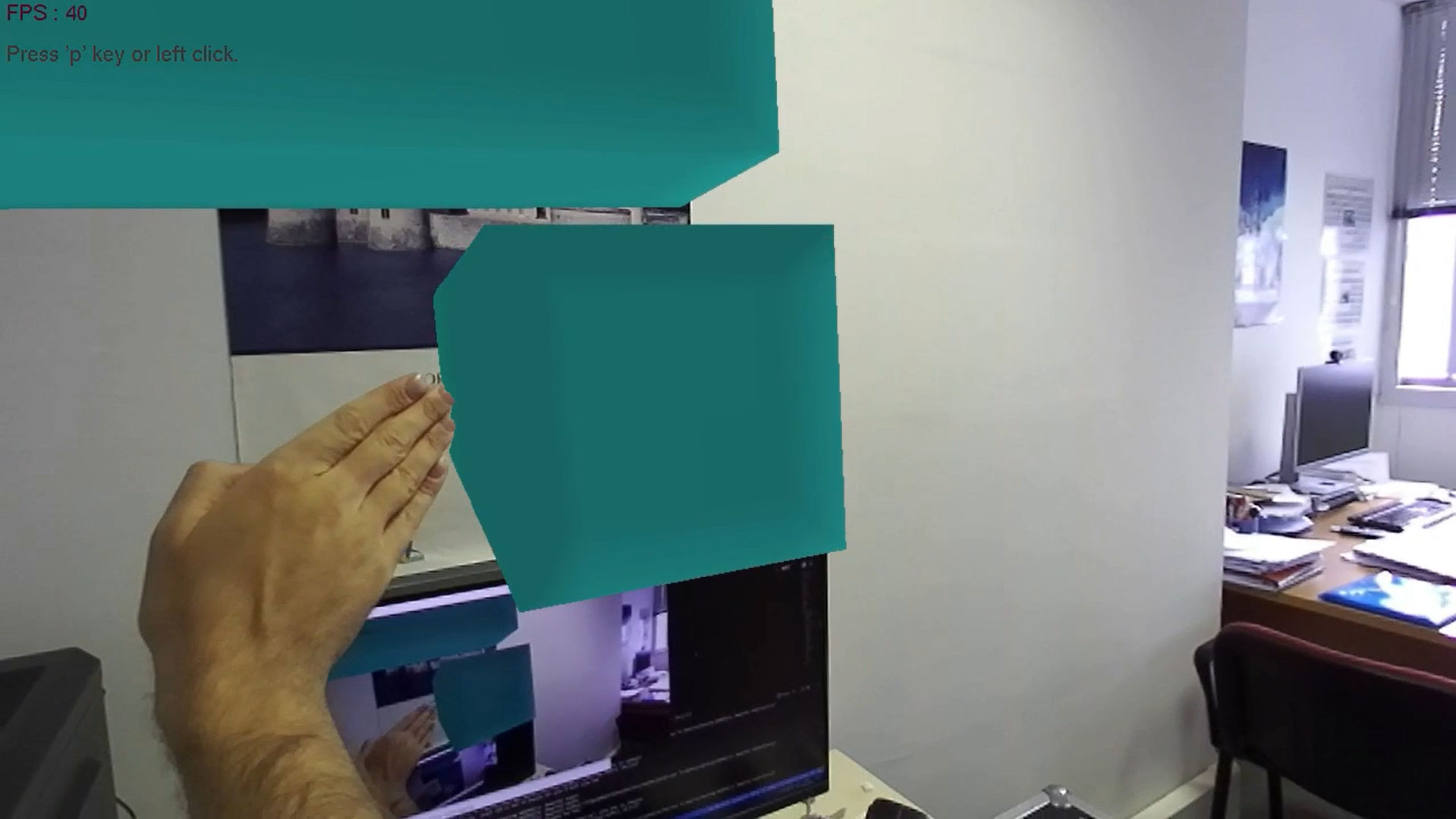}
  \end{subfigure}
\caption{Frames extracted from the interacting beams sequence. Color encodes the x-x stress field component associated with the displacement imposed by contact with a real object.}
\label{fig:beams_video}
\end{figure}

\myfigref{fig:beams_video} shows a real-time video sequence generated using the presented algorithm. The interaction of both the real objects (finger tips) and the virtual objects are simulated smoothly at more than 30 frames per second. It is important to highlight that during a video sequence the trained neural network is able to generalize to previously unseen configurations. The quantitative errors by the neural network in the real-time rollout predictions are shown in \myfigref{fig:boxplot}a, which remain below $1\%$ in position and velocity and $10\%$ in the stress tensor field.

The computational cost of the high fidelity FEM simulations used for the training of the neural network is $15$ s  per simulation, up to $795$ s for the whole dataset, and $243$ Mb in memory storage. Conversely, the training time of the neural network is $4.5$ h and the mean evaluation time is $9$ ms per snapshot with a memory storage of $12.3$ Mb. Thus, the use of a deep learning approach allows a drastic reduction of the online computation time with the inconvenient of larger offline training time. It is also worth noting that the network parameters are more than $10$ times lighter in terms of memory storage.

\subsection{Stanford bunny}\label{sec:bunny}

\begin{figure}[h]
\centering
\includegraphics[width=.4\textwidth]{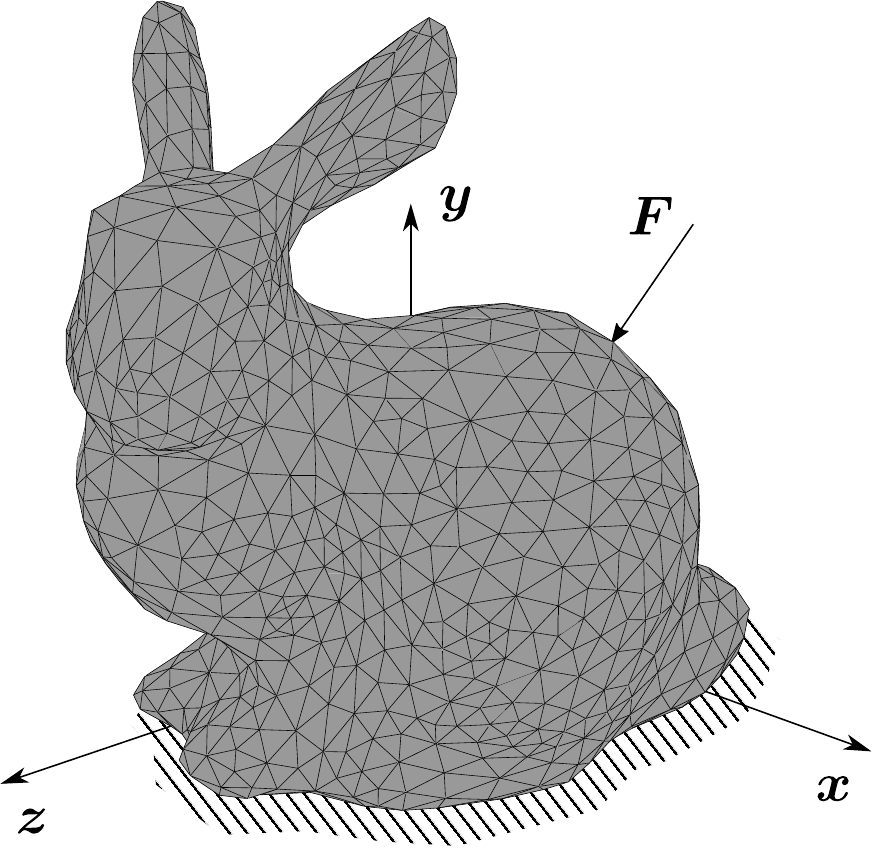}
\caption{Bunny mesh geometry with a concentrated application force and bottom encastre.\label{fig:bunny}}
\end{figure}

The second example is a bunny mesh from the Stanford 3D scanning repository, which is a more complex geometry as the one shown in the previous example. It is a standard reference model in computer graphics research. The model is represented in Fig. \ref{fig:bunny}.

The finite element mesh from which data are obtained consisted of $N_e=4941$ tetrahedral linear elements and $N=1352$ nodes. The constitutive parameters of the hyperelastic strain energy potential are $C_{10}=2.6\nexp{-1}$ and $D_1=4.9\nexp{-2}$. Similarly as the previous case, a concentrated load of $F=1$ is applied in 100 different positions with a perpendicular direction to the solid surface. The body is fixed to the ground plane by disabling displacements and rotations at the lower model nodes. Each simulation is composed of $N_T=20$ time increments of $\Delta t=5\nexp{-2}$ s. 

The graph neural network vertex and edge MLPs have two layers of $F_h=100$ neurons each, with 10 message passing sequential blocks. The training was performed for $N_\text{epoch}=1800$ epochs and learning rate $l_r=10^{-4}$.

\begin{figure}[h]
  \begin{subfigure}[t]{.49\textwidth}
    \centering
    \includegraphics[width=\textwidth]{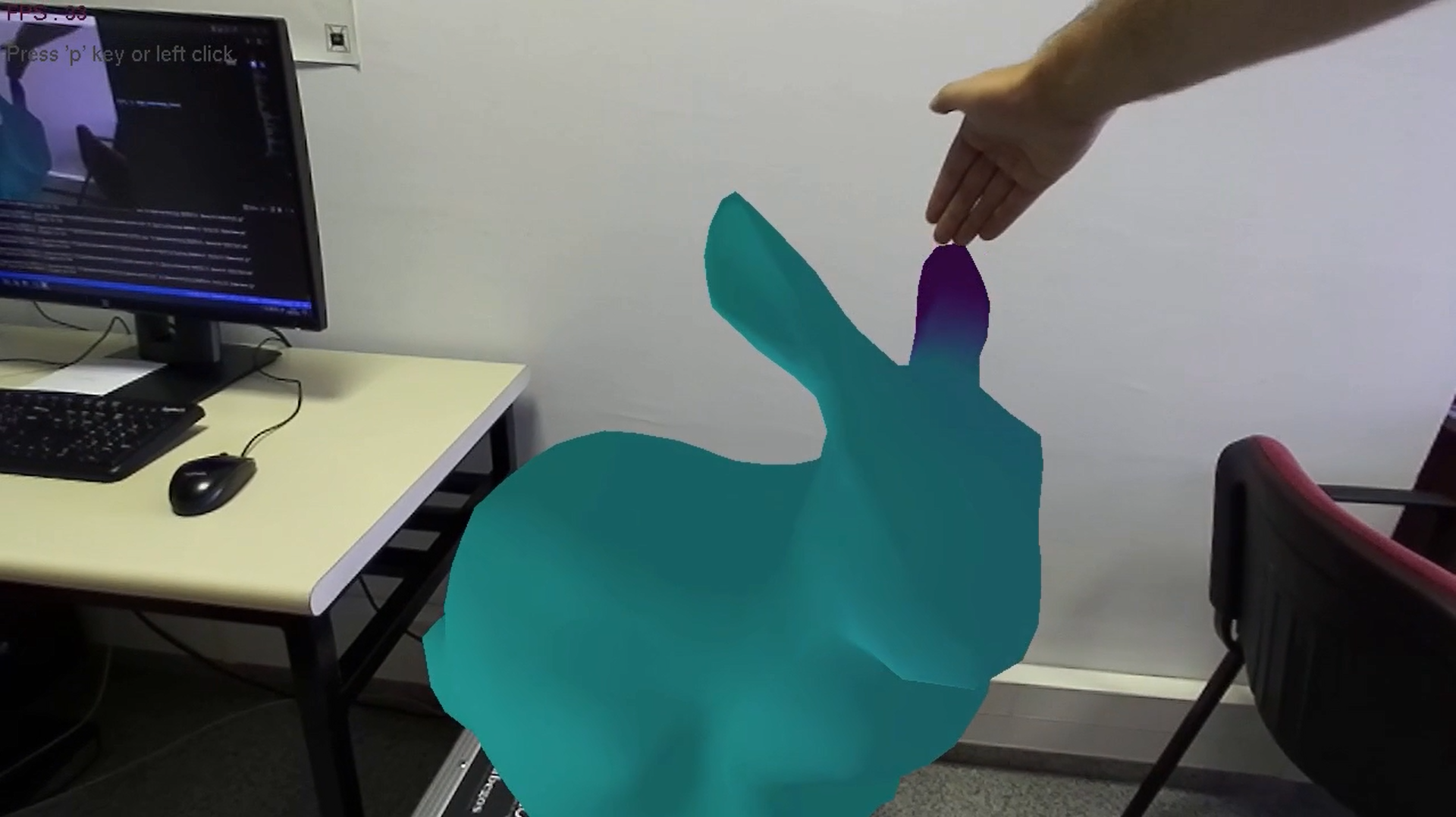}
  \end{subfigure}
  \hfill
  \begin{subfigure}[t]{.49\textwidth}
    \centering
    \includegraphics[width=\textwidth]{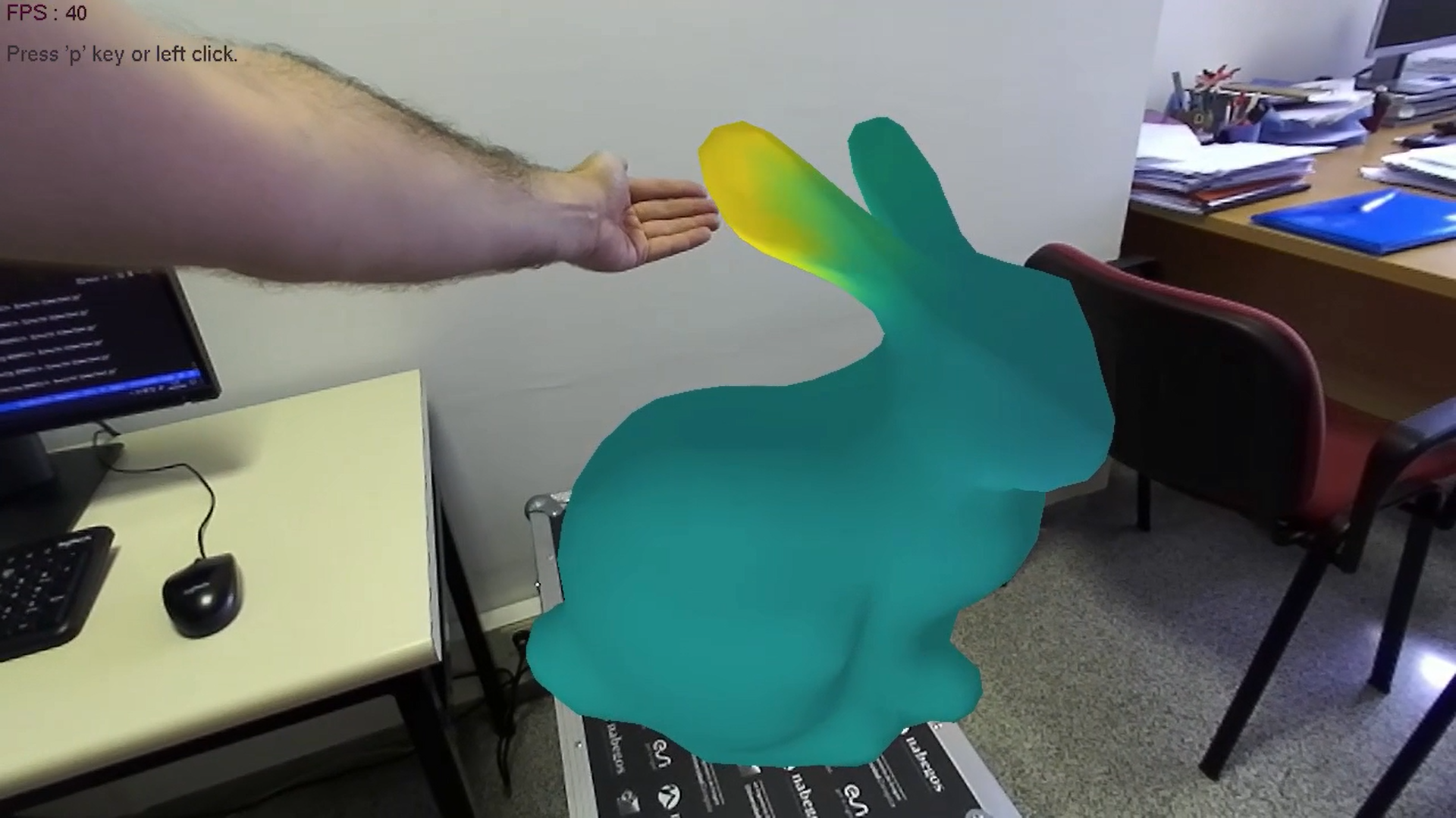}
  \end{subfigure}

  \medskip

  \begin{subfigure}[t]{.49\textwidth}
    \centering
    \includegraphics[width=\textwidth]{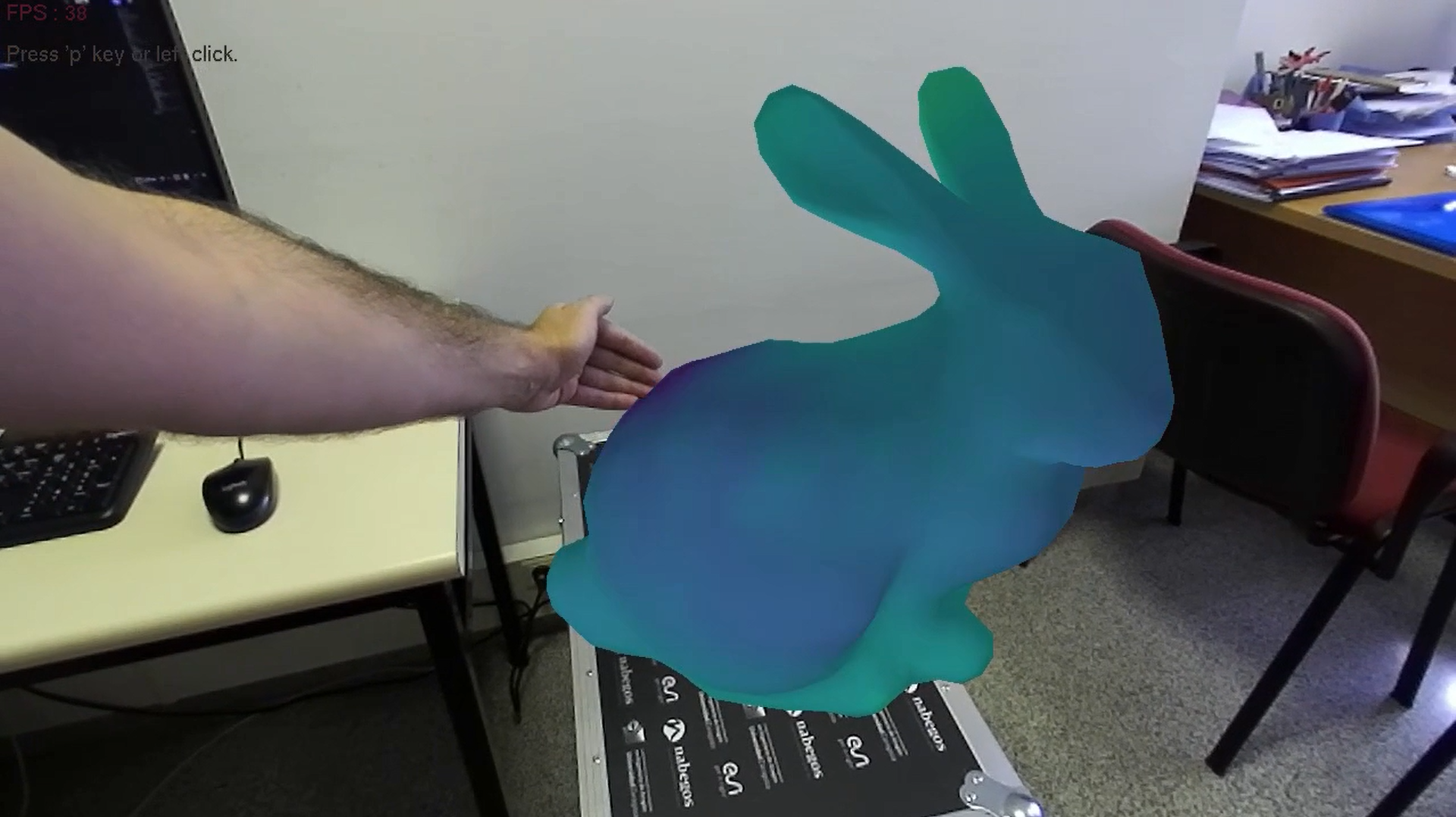}
  \end{subfigure}
  \hfill
  \begin{subfigure}[t]{.49\textwidth}
    \centering
    \includegraphics[width=\textwidth]{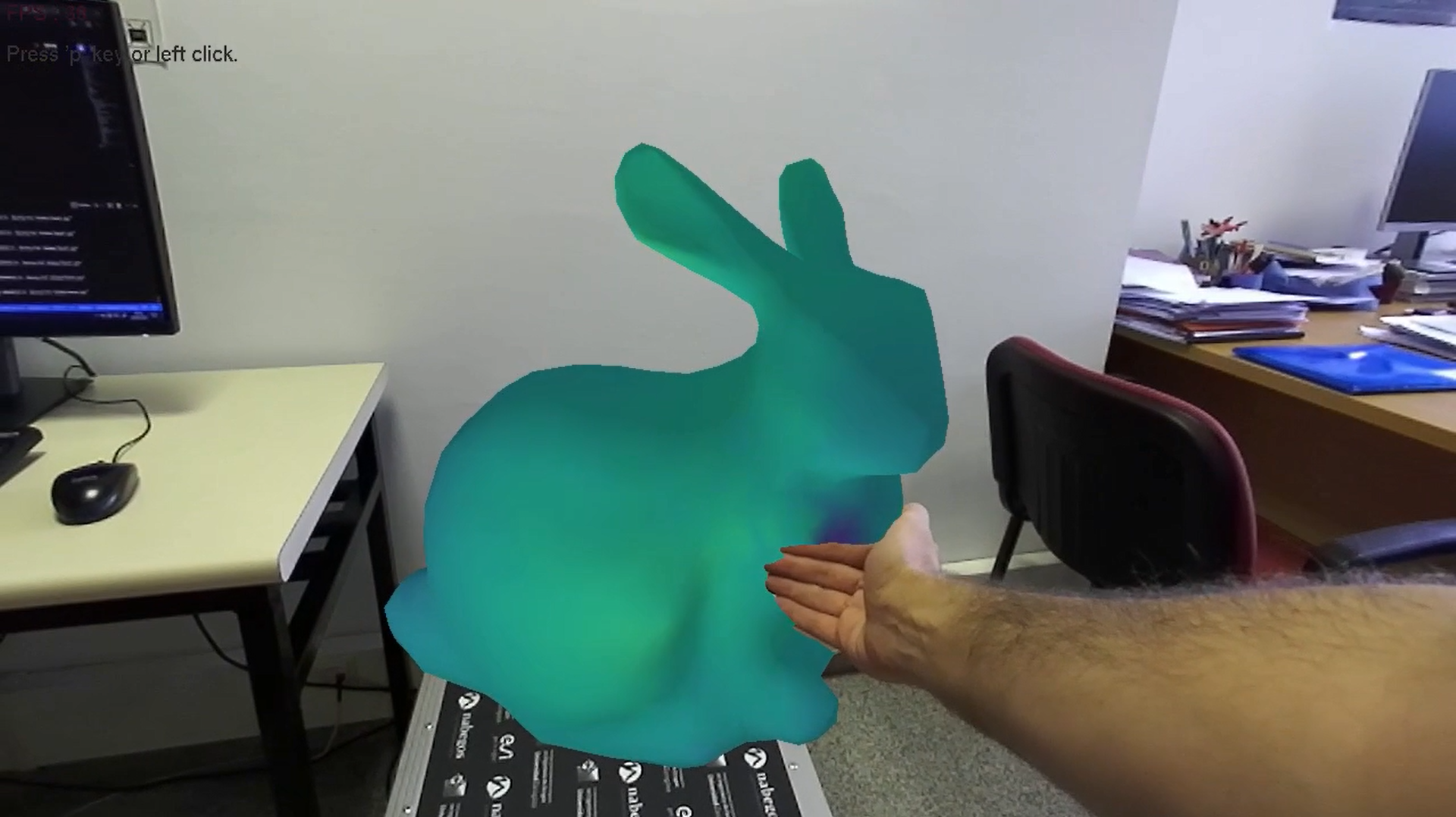}
  \end{subfigure}
\caption{Frames extracted from the bunny sequence. Color encode the x displacement field component imposed by contact with a real object.}
\label{fig:bunny_video}
\end{figure}

\begin{figure}[h]
\centering
\input{results/boxplot/boxplot.tex}
\caption{Box plots for the relative L2 error for all the rollout snapshots of the bunny dataset in both train and test cases. The state variables represented are position ($\bs{q}$), velocity ($\bs{v}$), and stress tensor ($\bs{\sigma}$).}
\label{fig:boxplot}
\end{figure}
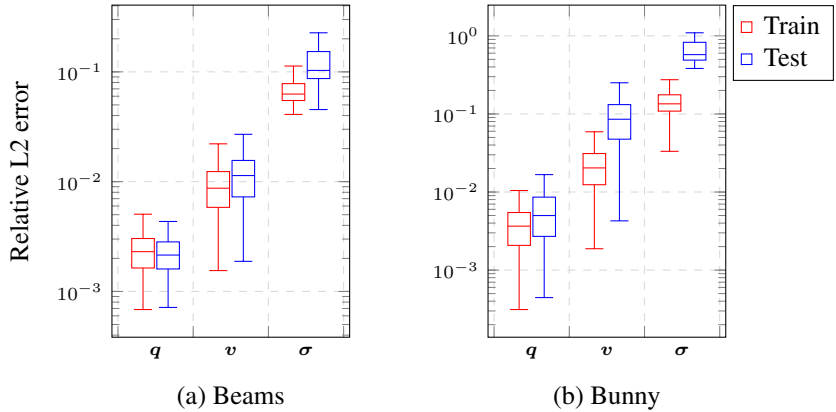

\myfigref{fig:bunny_video} shows a real-time video sequence generated using the bunny model, also with a minimum framerate of 30 frames per second. In this case, the stress field errors reported in \myfigref{fig:boxplot}b are higher due to the stress peaks at the single-node force application, which causes that the elasticity phenomena remain very local in space.

The computational cost of the high fidelity FEM simulations is $27$ s  per simulation, up to $2700$ s for the whole dataset and $833$ Mb in memory storage.  The dataset and edge count is much larger in this example, which increases the training time of the neural network to $20$ h and the evaluation time to $11$ ms per snapshot with the same memory storage as the previous example. Thus, the data compression is even bigger in this case. To address the high training time and errors due to stress peaks, several future improvements are discussed in the next section.

\section{Conclusions}\label{sec:conc}

We presented a real-time augmented reality simulator, which enables a user to interact with virtual deformable solids. The predictions are computed using the GENERIC structure of the system learnt with a message passing graph neural network. The enforcement of such physics constraints guarantees the fulfilment of the first and second laws of thermodynamics. The resulting algorithm has a wide variety of applications not only in the entertainment industry, but also in engineering design or manufacturing, where the visualization of augmented data superimposed in a real or virtual object might redefine the next generation of industry 4.0 and digital twins.

Our method has several limitations which might be addressed as future work. The scalability to bigger meshes is a challenging task, as graph neural networks can suffer from over-squashing or bottlenecks \cite{alon2020bottleneck, topping2021understanding}. The use of graph reduction techniques via graph autoencoders \cite{kipf2016variational} or U-nets \cite{gao2019graph} configurations can significantly reduce the computational requirements in larger meshes. This would also reduce the stress peaks discontinuities due to concentrated forces. The message passing algorithm itself is not able to handle domains where the boundary conditions are very far away from the prescribed forces. This can be mitigated by implementing a global attention vector so that certain dynamical information is reached to all the nodes of the domain instantly.

The visualization and data acquisition are possible due to the graphics acceleration. Even if the current work was implemented in a desktop computer as a proof of concept, only a small fraction of all the computational resources were used, so it can be extended to AR/VR headsets or to modern mobile devices. For the same reason, higher framerates might be achieved by fine-tuning and optimizing the proposed network structure or porting the code to a higher-performance language such as C++. Occlusions in real-time augmented reality software still remain as an open problem in the computer vision community \cite{parger2021unoc,yan2022augmented}. The stereo depth estimation is an approximation which might cause image artifacts in singular camera poses. For instance, some works handle the problem by using deep learning \cite{martin2018lookingood} but it is out of the scope of this manuscript.

This work is just a small step forward towards the new immersive technologies which can potentially deeply change our society. It is a multidisciplinary problem where computer vision, computer graphics, machine learning and computational mechanics must meet to define new algorithms for a new digital revolution.

\section*{Acknowledgements}

This material is based upon work supported in part by the Army Research Laboratory and the Army Research Office under contract/grant number W911NF2210271.
This work has also been partially funded by the Spanish Ministry of Science and Innovation, AEI /10.13039/501100011033, through Grant number PID2020-113463RB-C31, and by the \emph{Primeros Proyectos} Grant from Polytechnic University of Madrid, ETSII-UPM22-PM01. 

The support of ESI Group through the Chairs at ENSAM Paris, the DesCartes programme under its Campus for Research Excellence and Technological Enterprise (CREATE) programme and Universidad de Zaragoza is also gratefully acknowledged.

\bibliography{AR-GENERIC-ARXIV}
\bibliographystyle{unsrt}

\end{document}

%% file: header.tex
\newcommand{\bs}[1]{\boldsymbol{#1}}
\newcommand{\nexp}[1]{\cdot 10^{#1}}
\newcommand{\dpar}[2]{\frac{\partial #1}{\partial #2}}

\newcommand{\mysecref}[1]{Section~\ref{#1}}
\newcommand{\myeqref}[1]{Eq.~(\ref{#1})}
\newcommand{\myfigref}[1]{Fig.~\ref{#1}}

\usepackage{url}
\usepackage{pgfplots}
\usepackage{algorithm,algpseudocode}
\usepackage{algorithmicx}
\usepackage{amsmath}
\usepackage{amsfonts}
\usepackage{gensymb}
\usepackage{siunitx}\usepackage[colorlinks,bookmarksnumbered,bookmarksopen,citecolor=blue,urlcolor=blue,linkcolor=blue,]{hyperref}
\usepackage{xcolor}
\usepackage{subcaption}
\usepackage{graphicx}

\usepackage{pgfplotstable}
\usepgfplotslibrary{statistics}
\pgfplotsset{
/pgfplots/custom legend/.style={
legend image code/.code={
\draw [only marks,mark=square]
plot coordinates {(0.3cm,0cm)};
}, },}

\makeatletter
\pgfplotsset{
    boxplot prepared from table/.code={
        \def\tikz@plot@handler{\pgfplotsplothandlerboxplotprepared}%
        \pgfplotsset{
            /pgfplots/boxplot prepared from table/.cd,
            #1,
        }
    },
    /pgfplots/boxplot prepared from table/.cd,
        table/.code={\pgfplotstablecopy{#1}\to\boxplot@datatable},
        row/.initial=0,
        make style readable from table/.style={
            #1/.code={
                \pgfplotstablegetelem{\pgfkeysvalueof{/pgfplots/boxplot prepared from table/row}}{##1}\of\boxplot@datatable
                \pgfplotsset{boxplot/#1/.expand once={\pgfplotsretval}}
            }
        },
        make style readable from table=lower whisker,
        make style readable from table=upper whisker,
        make style readable from table=lower quartile,
        make style readable from table=upper quartile,
        make style readable from table=median,
        make style readable from table=lower notch,
        make style readable from table=upper notch
}
\makeatother

%% file: results/boxplot/boxplot.tex
\pgfplotstableread{results/beams_box/error_beams_train.txt}\errorbeamstrain
\pgfplotstableread{results/beams_box/error_beams_test.txt}\errorbeamstest
\pgfplotstableread{results/bunny_box/error_bunny_train.txt}\errorbunnytrain
\pgfplotstableread{results/bunny_box/error_bunny_test.txt}\errorbunnytest

\begin{tikzpicture}
\pgfplotsset{width=\textwidth, height=6cm}

  \begin{semilogyaxis}
  [xshift=0cm,xlabel=(a) Beams,
  area legend, boxplot/draw direction=y,
  grid=major, 
  grid style={dashed,gray!30}, 
  cycle list={{red},{blue}},
  boxplot={draw position={1/3 + floor(\plotnumofactualtype/2) + 1/3*mod(\plotnumofactualtype,2)},box extend=0.3,},
  x=1cm,xtick={0,1,2,...,10},x tick label as interval,
  xticklabels={{$\bs{q}$},{$\bs{v}$},{$\bs{\sigma}$}},
  ticklabel style={font=\scriptsize},
  ylabel={Relative L2 error},
  custom legend,legend pos=outer north east,legend cell align=left,
  ] 
	\addplot+[boxplot prepared from table={table=\errorbeamstrain,row=0,
    lower whisker=lw,
    upper whisker=uw,
    lower quartile=lq,
    upper quartile=uq,
    median=med}, boxplot prepared]
    coordinates {};
	\addplot+[boxplot prepared from table={table=\errorbeamstest,row=0,
    lower whisker=lw,
    upper whisker=uw,
    lower quartile=lq,
    upper quartile=uq,
    median=med}, boxplot prepared] 
    coordinates {}; 
    
	\addplot+[boxplot prepared from table={table=\errorbeamstrain,row=1,
    lower whisker=lw,
    upper whisker=uw,
    lower quartile=lq,
    upper quartile=uq,
    median=med}, boxplot prepared] 
    coordinates {}; 
	\addplot+[boxplot prepared from table={table=\errorbeamstest,row=1,
    lower whisker=lw,
    upper whisker=uw,
    lower quartile=lq,
    upper quartile=uq,
    median=med}, boxplot prepared] 
    coordinates {}; 

	\addplot+[boxplot prepared from table={table=\errorbeamstrain,row=2,
    lower whisker=lw,
    upper whisker=uw,
    lower quartile=lq,
    upper quartile=uq,
    median=med}, boxplot prepared] 
    coordinates {}; 
	\addplot+[boxplot prepared from table={table=\errorbeamstest,row=2,
    lower whisker=lw,
    upper whisker=uw,
    lower quartile=lq,
    upper quartile=uq,
    median=med}, boxplot prepared] 
    coordinates {}; 

  \end{semilogyaxis}

  \begin{semilogyaxis}
  [xshift=5cm,xlabel=(b) Bunny,
  area legend, boxplot/draw direction=y,
  grid=major, 
  grid style={dashed,gray!30}, 
  cycle list={{red},{blue}},
  boxplot={draw position={1/3 + floor(\plotnumofactualtype/2) + 1/3*mod(\plotnumofactualtype,2)},box extend=0.3,},
  x=1cm,xtick={0,1,2,...,10},x tick label as interval,
  xticklabels={{$\bs{q}$},{$\bs{v}$},{$\bs{\sigma}$}},
  ticklabel style={font=\scriptsize},
  custom legend,legend pos=outer north east,legend cell align=left,
  legend entries = {Train, Test},
  ] 
	\addplot+[boxplot prepared from table={table=\errorbunnytrain,row=0,
    lower whisker=lw,
    upper whisker=uw,
    lower quartile=lq,
    upper quartile=uq,
    median=med}, boxplot prepared]
    coordinates {};
	\addplot+[boxplot prepared from table={table=\errorbunnytest,row=0,
    lower whisker=lw,
    upper whisker=uw,
    lower quartile=lq,
    upper quartile=uq,
    median=med}, boxplot prepared] 
    coordinates {}; 
    
	\addplot+[boxplot prepared from table={table=\errorbunnytrain,row=1,
    lower whisker=lw,
    upper whisker=uw,
    lower quartile=lq,
    upper quartile=uq,
    median=med}, boxplot prepared] 
    coordinates {}; 
	\addplot+[boxplot prepared from table={table=\errorbunnytest,row=1,
    lower whisker=lw,
    upper whisker=uw,
    lower quartile=lq,
    upper quartile=uq,
    median=med}, boxplot prepared] 
    coordinates {}; 

	\addplot+[boxplot prepared from table={table=\errorbunnytrain,row=2,
    lower whisker=lw,
    upper whisker=uw,
    lower quartile=lq,
    upper quartile=uq,
    median=med}, boxplot prepared] 
    coordinates {}; 
    \addlegendentry{~Train}
	\addplot+[boxplot prepared from table={table=\errorbunnytest,row=2,
    lower whisker=lw,
    upper whisker=uw,
    lower quartile=lq,
    upper quartile=uq,
    median=med}, boxplot prepared] 
    coordinates {}; 
    \addlegendentry{~Test}

  \end{semilogyaxis}
  
\end{tikzpicture}

%% file: AR-GENERIC-ARXIV.bbl
\begin{thebibliography}{10}

\bibitem{allam2022metaverse}
Zaheer Allam, Ayyoob Sharifi, Simon~Elias Bibri, David~Sydney Jones, and John
  Krogstie.
\newblock The metaverse as a virtual form of smart cities: opportunities and
  challenges for environmental, economic, and social sustainability in urban
  futures.
\newblock {\em Smart Cities}, 5(3):771--801, 2022.

\bibitem{veeraiah2022enhancement}
Vivek Veeraiah, P~Gangavathi, Shahanawai Ahamad, Suryansh~Bhaskar Talukdar,
  Ankur Gupta, and Veera Talukdar.
\newblock Enhancement of meta verse capabilities by iot integration.
\newblock In {\em 2022 2nd International Conference on Advance Computing and
  Innovative Technologies in Engineering (ICACITE)}, pages 1493--1498. IEEE,
  2022.

\bibitem{rospigliosi2022metaverse}
Pericles~‘asher’ Rospigliosi.
\newblock Metaverse or simulacra? roblox, minecraft, meta and the turn to
  virtual reality for education, socialisation and work, 2022.

\bibitem{wang2022metasocieties}
Fei-Yue Wang, Rui Qin, Xiao Wang, and Bin Hu.
\newblock Metasocieties in metaverse: Metaeconomics and metamanagement for
  metaenterprises and metacities.
\newblock {\em IEEE Transactions on Computational Social Systems}, 9(1):2--7,
  2022.

\bibitem{mystakidis2022metaverse}
Stylianos Mystakidis.
\newblock Metaverse.
\newblock {\em Encyclopedia}, 2(1):486--497, 2022.

\bibitem{kraus2022facebook}
Sascha Kraus, Dominik~K Kanbach, Peter~M Krysta, Maurice~M Steinhoff, and Nino
  Tomini.
\newblock Facebook and the creation of the metaverse: radical business model
  innovation or incremental transformation?
\newblock {\em International Journal of Entrepreneurial Behavior \& Research},
  2022.

\bibitem{hummel2019leveraging}
Mathias Hummel and Kees~van Kooten.
\newblock Leveraging nvidia omniverse for in situ visualization.
\newblock In {\em International Conference on High Performance Computing},
  pages 634--642. Springer, 2019.

\bibitem{li2022exploring}
Xiao Li, Baris~Can Yalcin, Olga-Orsalia Christidi-Loumpasefski, Carol
  Martinez~Luna, Maxime Hubert~Delisle, Gonzalo Rodriguez, James Zheng, and
  Miguel~Angel Olivares~Mendez.
\newblock Exploring nvidia omniverse for future space resources missions.
\newblock 2022.

\bibitem{maciel2009using}
Anderson Maciel, Tansel Halic, Zhonghua Lu, Luciana~P Nedel, and Suvranu De.
\newblock Using the physx engine for physics-based virtual surgery with force
  feedback.
\newblock {\em The International Journal of Medical Robotics and Computer
  Assisted Surgery}, 5(3):341--353, 2009.

\bibitem{d2013physx}
Antonio D'Andrea, Monica Reggiani, Andrea Turolla, Davide Cattin, and Roberto
  Oboe.
\newblock A physx-based framework to develop rehabilitation using haptic and
  virtual reality.
\newblock In {\em 2013 IEEE International Symposium on Industrial Electronics},
  pages 1--6. IEEE, 2013.

\bibitem{wang2015method}
Lei Wang, Zhenwen Wang, and Hua Xu.
\newblock A method for 3d rock fracturing simulation based havok.
\newblock In {\em 2015 6th IEEE International Conference on Software
  Engineering and Service Science (ICSESS)}, pages 898--901. IEEE, 2015.

\bibitem{berkooz1993proper}
Gal Berkooz, Philip Holmes, and John~L Lumley.
\newblock The proper orthogonal decomposition in the analysis of turbulent
  flows.
\newblock {\em Annual review of fluid mechanics}, 25(1):539--575, 1993.

\bibitem{rama2016real}
Ritesh~R Rama, Sebastian Skatulla, and Carlo Sansour.
\newblock Real-time modelling of diastolic filling of the heart using the
  proper orthogonal decomposition with interpolation.
\newblock {\em International Journal of Solids and Structures}, 96:409--422,
  2016.

\bibitem{prud2002reliable}
Christophe Prud’Homme, Dimitrios~V Rovas, Karen Veroy, Luc Machiels, Yvon
  Maday, Anthony~T Patera, and Gabriel Turinici.
\newblock Reliable real-time solution of parametrized partial differential
  equations: Reduced-basis output bound methods.
\newblock {\em J. Fluids Eng.}, 124(1):70--80, 2002.

\bibitem{manzoni2015reduced}
Andrea Manzoni, Filippo Salmoiraghi, and Luca Heltai.
\newblock Reduced basis isogeometric methods (rb-iga) for the real-time
  simulation of potential flows about parametrized naca airfoils.
\newblock {\em Computer Methods in Applied Mechanics and Engineering},
  284:1147--1180, 2015.

\bibitem{scholkopf1997kernel}
Bernhard Sch{\"o}lkopf, Alexander Smola, and Klaus-Robert M{\"u}ller.
\newblock Kernel principal component analysis.
\newblock In {\em International conference on artificial neural networks},
  pages 583--588. Springer, 1997.

\bibitem{roweis2000nonlinear}
Sam~T Roweis and Lawrence~K Saul.
\newblock Nonlinear dimensionality reduction by locally linear embedding.
\newblock {\em science}, 290(5500):2323--2326, 2000.

\bibitem{moya2019learning}
Beatriz Moya, David Gonz{\'a}lez, Ic{\'\i}ar Alfaro, Francisco Chinesta, and
  Elias Cueto.
\newblock Learning slosh dynamics by means of data.
\newblock {\em Computational Mechanics}, 64(2):511--523, 2019.

\bibitem{badias2017local}
Alberto Bad{\'\i}as, David Gonz{\'a}lez, Iciar Alfaro, Francisco Chinesta, and
  Elias Cueto.
\newblock Local proper generalized decomposition.
\newblock {\em International Journal for Numerical Methods in Engineering},
  112(12):1715--1732, 2017.

\bibitem{badias2020real}
Alberto Bad{\'\i}as, David Gonz{\'a}lez, Ic{\'\i}ar Alfaro, Francisco Chinesta,
  and El{\'\i}as Cueto.
\newblock Real-time interaction of virtual and physical objects in mixed
  reality applications.
\newblock {\em International Journal for Numerical Methods in Engineering},
  121(17):3849--3868, 2020.

\bibitem{fulton2019latent}
Lawson Fulton, Vismay Modi, David Duvenaud, David~IW Levin, and Alec Jacobson.
\newblock Latent-space dynamics for reduced deformable simulation.
\newblock In {\em Computer graphics forum}, volume~38, pages 379--391. Wiley
  Online Library, 2019.

\bibitem{chen2022crom}
Peter~Yichen Chen, Jinxu Xiang, Dong~Heon Cho, Yue Chang, GA~Pershing,
  Henrique~Teles Maia, Maurizio Chiaramonte, Kevin Carlberg, and Eitan
  Grinspun.
\newblock Crom: Continuous reduced-order modeling of pdes using implicit neural
  representations.
\newblock {\em arXiv preprint arXiv:2206.02607}, 2022.

\bibitem{fresca2022deep}
Stefania Fresca, Giorgio Gobat, Patrick Fedeli, Attilio Frangi, and Andrea
  Manzoni.
\newblock Deep learning-based reduced order models for the real-time simulation
  of the nonlinear dynamics of microstructures.
\newblock {\em International Journal for Numerical Methods in Engineering},
  123(20):4749--4777, 2022.

\bibitem{odot2022deepphysics}
Alban Odot, Ryadh Haferssas, and St{\'e}phane Cotin.
\newblock Deepphysics: A physics aware deep learning framework for real-time
  simulation.
\newblock {\em International Journal for Numerical Methods in Engineering},
  123(10):2381--2398, 2022.

\bibitem{romero2022embodied}
Javier Romero, Dimitrios Tzionas, and Michael~J Black.
\newblock Embodied hands: Modeling and capturing hands and bodies together.
\newblock {\em arXiv preprint arXiv:2201.02610}, 2022.

\bibitem{romero2022contact}
Cristian Romero, Dan Casas, Maurizio~M Chiaramonte, and Miguel~A Otaduy.
\newblock Contact-centric deformation learning.
\newblock {\em ACM Transactions on Graphics (TOG)}, 41(4):1--11, 2022.

\bibitem{raissi2018hidden}
Maziar Raissi and George~Em Karniadakis.
\newblock Hidden physics models: Machine learning of nonlinear partial
  differential equations.
\newblock {\em Journal of Computational Physics}, 357:125--141, 2018.

\bibitem{raissi2019physics}
Maziar Raissi, Paris Perdikaris, and George~E Karniadakis.
\newblock Physics-informed neural networks: A deep learning framework for
  solving forward and inverse problems involving nonlinear partial differential
  equations.
\newblock {\em Journal of Computational Physics}, 378:686--707, 2019.

\bibitem{eivazi2022physics}
Hamidreza Eivazi, Mojtaba Tahani, Philipp Schlatter, and Ricardo Vinuesa.
\newblock Physics-informed neural networks for solving reynolds-averaged
  navier--stokes equations.
\newblock {\em Physics of Fluids}, 34(7):075117, 2022.

\bibitem{eivazi2022bphysics}
Hamidreza Eivazi and Ricardo Vinuesa.
\newblock Physics-informed deep-learning applications to experimental fluid
  mechanics.
\newblock {\em arXiv preprint arXiv:2203.15402}, 2022.

\bibitem{cai2022physics}
Shengze Cai, Zhiping Mao, Zhicheng Wang, Minglang Yin, and George~Em
  Karniadakis.
\newblock Physics-informed neural networks (pinns) for fluid mechanics: A
  review.
\newblock {\em Acta Mechanica Sinica}, pages 1--12, 2022.

\bibitem{sanchez2020learning}
Alvaro Sanchez-Gonzalez, Jonathan Godwin, Tobias Pfaff, Rex Ying, Jure
  Leskovec, and Peter Battaglia.
\newblock Learning to simulate complex physics with graph networks.
\newblock In {\em International Conference on Machine Learning}, pages
  8459--8468. PMLR, 2020.

\bibitem{sanchez2019hamiltonian}
Alvaro Sanchez-Gonzalez, Victor Bapst, Kyle Cranmer, and Peter Battaglia.
\newblock Hamiltonian graph networks with ode integrators.
\newblock {\em arXiv preprint arXiv:1909.12790}, 2019.

\bibitem{chen2019symplectic}
Zhengdao Chen, Jianyu Zhang, Martin Arjovsky, and L{\'e}on Bottou.
\newblock Symplectic recurrent neural networks.
\newblock {\em arXiv preprint arXiv:1909.13334}, 2019.

\bibitem{hernandez2021structure}
Quercus Hern{\'a}ndez, Alberto Bad{\'\i}as, David Gonz{\'a}lez, Francisco
  Chinesta, and El{\'\i}as Cueto.
\newblock Structure-preserving neural networks.
\newblock {\em Journal of Computational Physics}, 426:109950, 2021.

\bibitem{hernandez2022thermodynamics}
Quercus Hern{\'a}ndez, Alberto Bad{\'\i}as, Francisco Chinesta, and El{\'\i}as
  Cueto.
\newblock Thermodynamics-informed graph neural networks.
\newblock {\em arXiv preprint arXiv:2203.01874}, 2022.

\bibitem{battaglia2018relational}
Peter~W Battaglia, Jessica~B Hamrick, Victor Bapst, Alvaro Sanchez-Gonzalez,
  Vinicius Zambaldi, Mateusz Malinowski, Andrea Tacchetti, David Raposo, Adam
  Santoro, Ryan Faulkner, et~al.
\newblock Relational inductive biases, deep learning, and graph networks.
\newblock {\em arXiv preprint arXiv:1806.01261}, 2018.

\bibitem{ottinger1997dynamics}
Hans~Christian {\"O}ttinger and Miroslav Grmela.
\newblock Dynamics and thermodynamics of complex fluids. ii. illustrations of a
  general formalism.
\newblock {\em Physical Review E}, 56(6):6633, 1997.

\bibitem{grmela1997dynamics}
Miroslav Grmela and Hans~Christian {\"O}ttinger.
\newblock Dynamics and thermodynamics of complex fluids. i. development of a
  general formalism.
\newblock {\em Physical Review E}, 56(6):6620, 1997.

\bibitem{zhang2020mediapipe}
Fan Zhang, Valentin Bazarevsky, Andrey Vakunov, Andrei Tkachenka, George Sung,
  Chuo-Ling Chang, and Matthias Grundmann.
\newblock Mediapipe hands: On-device real-time hand tracking.
\newblock {\em arXiv preprint arXiv:2006.10214}, 2020.

\bibitem{fey2019fast}
Matthias Fey and Jan~Eric Lenssen.
\newblock Fast graph representation learning with pytorch geometric.
\newblock {\em arXiv preprint arXiv:1903.02428}, 2019.

\bibitem{alon2020bottleneck}
Uri Alon and Eran Yahav.
\newblock On the bottleneck of graph neural networks and its practical
  implications.
\newblock {\em arXiv preprint arXiv:2006.05205}, 2020.

\bibitem{topping2021understanding}
Jake Topping, Francesco Di~Giovanni, Benjamin~Paul Chamberlain, Xiaowen Dong,
  and Michael~M Bronstein.
\newblock Understanding over-squashing and bottlenecks on graphs via curvature.
\newblock {\em arXiv preprint arXiv:2111.14522}, 2021.

\bibitem{kipf2016variational}
Thomas~N Kipf and Max Welling.
\newblock Variational graph auto-encoders.
\newblock {\em arXiv preprint arXiv:1611.07308}, 2016.

\bibitem{gao2019graph}
Hongyang Gao and Shuiwang Ji.
\newblock Graph u-nets.
\newblock In {\em international conference on machine learning}, pages
  2083--2092. PMLR, 2019.

\bibitem{parger2021unoc}
Mathias Parger, Chengcheng Tang, Yuanlu Xu, Christopher~David Twigg, Lingling
  Tao, Yijing Li, Robert Wang, and Markus Steinberger.
\newblock Unoc: Understanding occlusion for embodied presence in virtual
  reality.
\newblock {\em IEEE Transactions on Visualization and Computer Graphics}, 2021.

\bibitem{yan2022augmented}
Wei Yan.
\newblock Augmented reality instructions for construction toys enabled by
  accurate model registration and realistic object/hand occlusions.
\newblock {\em Virtual Reality}, 26(2):465--478, 2022.

\bibitem{martin2018lookingood}
Ricardo Martin-Brualla, Rohit Pandey, Shuoran Yang, Pavel Pidlypenskyi,
  Jonathan Taylor, Julien Valentin, Sameh Khamis, Philip Davidson, Anastasia
  Tkach, Peter Lincoln, et~al.
\newblock Lookingood: Enhancing performance capture with real-time neural
  re-rendering.
\newblock {\em arXiv preprint arXiv:1811.05029}, 2018.

\end{thebibliography}
